\documentclass[usegraphicx,useAMS,usenatbib]{mn2e}

\title[Disc precession in Bo 158]{ Disc precession in the M31 dipping X-ray binary Bo 158?}
\author[R. Barnard et al.]{ R. Barnard,$^1$\thanks{Email address: R.Barnard@open.ac.uk} S. B. Foulkes,$^1$ C. A. Haswell,$^{1}$ U. Kolb,$^{1}$
\newauthor
 J. P. Osborne$^2$ and J. R. Murray$^3$\\
$^1$The Open University, Walton Hall, Milton Keynes, Buckinghamshire, MK7 6AA, UK\\
$^2$ The University of Leicester, University Road, Leicester, LE1 7RH, UK\\
$^3$ Swinburne University of Technology, PO Box 218, Hawthorne, Victoria 3122, Australia
}

\begin{document}
\date{Accepted  2005  November 7. Received 2005 February 16}
\pagerange{} \pubyear{}
\maketitle

\begin{abstract}
We present results from three  XMM-Newton observations of the M31 low mass X-ray binary XMMU\thinspace J004314.4+410726.3 (Bo 158), spaced over 3 days in 2004, July. Bo 158 was the  first dipping LMXB to be discovered in M31.   Periodic intensity dips were previously seen to occur on a 2.78-hr
period, due to absorption in material that is raised out of the plane of
the accretion disc. The report of these observations stated that the dip
depth was anti-correlated with source intensity. In light of the 2004 XMM-Newton observations of Bo 158,  we suggest that the dip variation is due to  precession of the accretion disc. This is to be expected in LMXBs with a mass ratio $\la$ 0.3
(period $\la$ 4 hr), as the disc reaches the 3:1 resonance with the binary
companion, causing elongation and precession of the disc. A smoothed
particle hydrodynamics simulation of the disc in this system shows retrograde  rotation of a disc warp on a period of $\sim$11 $P_{\rm orb}$, and prograde  disc precession on a period of 29$\pm$1 $P_{\rm orb}$. This is  consistent with the observed variation in the depth of the dips. We find that the dipping behaviour is most likely to be modified by the disc precession, hence we predict that the dipping behaviour repeats on a 81$\pm$3 hr cycle.
\end{abstract}


%\titlerunning{Disc precession in Bo 158}
%\maketitle

\section{Introduction}
\label{intro}

Bo 158 is source number 158 in the catalogue of globular clusters that were identified in M31 by \citet{bat87}. Its X-ray counterpart was discovered by the Einstein Observatory  \citep[ source number 81]{tf91}, and is located at  $\alpha$ = 00$^{\rm h}$43$^{\rm m}$14.2$^{\rm s}$, $\delta$ = 41$\degr$07$\arcmin$26.3$\arcsec$ \citep{dis02}. \citet{tru02} identified the X-ray source as a likely low mass X-ray binary (LMXB) with a neutron star primary; following their work, we will use the designation ``Bo 158'' to describe the X-ray source here.

\citet{tru02} report $\sim$83\% modulation in the 0.3--10 keV flux of Bo 158 on a 2.78 hour period during the $\sim$60 ks 2002 January XMM-Newton observation. The modulation resembles the intensity dips seen in high inclination LMXBs due to photo-electric absorption of X-rays by material that is raised above the body of the  accretion disc \citep{ws82}.
 The authors comment that the dipping is energy-independent, and discuss two possible mechanisms: obscuration of the central X-ray source by highly ionised material that scatters X-rays out of the line of sight, and the partial covering of an extended source by an opaque absorber.  They also report $\sim$30\% dips in the 2000, June XMM-Newton lightcurve  and $\sim$50\% dips in the 0.2--2.0 keV lightcurve of the 1991, June 26 ROSAT/PSPC observation. However, no significant dips were found in the 0.3--10 keV lightcurve of the 2001 June XMM-Newton observation; the authors placed a 2$\sigma$ upper limit of 10\% on the modulation. 

\citet{tru02} obtained fluxes in the 0.3--10 keV band for the three XMM-Newton observations by fitting absorbed Comptonisation models to the combined spectra of the EPIC-pn, MOS1 and MOS2, and concluded  that the depth of  the intensity modulation was anti-correlated with the source luminosity.  We present further XMM-Newton observations and modelling results which suggest that the variation in dipping behaviour may  instead be due to  precession in the accretion disc. Such behaviour is associated with the ``superhump'' phenomenon that is observed in interacting binaries where the mass ratio of the secondary to the primary is smaller than $\sim$0.3 \citep{wk91}. Superhumps are briefly reviewed in Sect.~\ref{shumps}, followed by details of the observations and data analysis in Sect.~\ref{obs}, and our results in Sect.~\ref{res}. Numerical modelling of the system is discussed in Sect. 5; the system was simulated by  a  3D smoothed particle hydrodynamics (SPH) code. We present our discussion in Sect.~\ref{discuss}, and finally our conclusion in Sect.~7.

\section{Superhumps}
\label{shumps}

Superhumps were first identified in the superoutbursts of the  SU UMa sub-class of cataclysmic variables. They are manifested as a periodic increase in the optical brightness on a period that is a few percent longer than the orbital period  \citep{vog74,war75}. SU UMas are a subclass of  dwarf novae with short orbital periods ($\la$2 hr) that exhibit particularly long, bright superoutbursts, separated by several outbursts that are typical of all dwarf novae; the superoutburst intervals are $\ga$5 times longer than the normal outbursts. \citep{vog80}.

 In the model proposed by \citet{osa89}, a superoutburst is occurs when a normal outburst is enhanced by a tidal instability; this occurs when the outer disc reaches a 3:1 resonance with the secondary.  
The disc is small at the start of the superoutburst cycle, well within the radius of tidal instability, and little angular momentum is removed by tidal interaction during the first normal outburst. Hence, the disc cannot accrete all of its mass onto the neutron star, and the size of the disc increases with successive outbursts, until the 3:1 resonance is reached \citep{osa89}.
The  additional tidal forces exerted on the disc by the secondary at this stage cause the disc to elongate and precess, and also greatly enhance the loss of angular momentum, so that the disc contracts, and most of the disc material is snow-ploughed  onto the neutron star, causing the superoutburst \citep{osa89}. 
The disc precession is prograde in the  rest frame, and the secondary repeats its motion with respect to the disc on the beat period between the orbital and precession periods, slightly longer than the orbital period. The secondary modulates the disc's viscous dissipation on this period, giving rise to maxima in the optical lightcurve, known as superhumps.

The requirement for the 3:1 resonance to fall within the disc's tidal radius is that the mass ratio of the secondary to the primary be less than  $\sim$0.33 \citep{wk91}. If we assume that the secondary is a main sequence star that fills its Roche lobe, then the relation $m_2$ $\simeq$ 0.11 $P_{\rm hr}$ holds, where $m_{2}$ is the mass of the secondary in solar units and $P_{\rm hr}$ is the orbital period in hours \citep*[e.g.][]{fkr92}. Hence any accreting binary that has a short enough orbital period may exhibit superhumps. Indeed,  there exists a class of short-period, persistently bright CVs  that exhibit permanent superhumps \citep{pat99,rn00}.

\citet{has01} discuss analogous superhump behaviour in LMXBs. Although superhumps have been found in the optical lightcurves of several black hole and neutron star LMXBs,
 they cannot be produced by the same mechanism, since the optical output of X-ray bright LMXBs is dominated by reprocessed X-rays. Instead, \citet{has01} proposed that the modulation is due to variation of the solid angle that the disc subtends to the X-ray source (on the superhump period); in this model, superhumping might be expected to be more prominent in low inclination systems.

4U\thinspace  1916$-$053 is a neutron star  LMXB, with an X-ray period of 50.00 min and an optical period of 50.4589 min \citep{cgc95}. \citet{has01} show that LMXBs with orbital periods shorter than $\sim$4.2 hr are likely to exhibit superhumps, and identified 4U\thinspace 1916$-$053 as a persistent irradiated  superhumping source.
 4U\thinspace 1916$-$053  is a high inclination system, with periodic intensity dips in the X-ray lightcurve \citep{ws82}. These dips are due to photo-electric absorption by material on the outer edge of the accretion disc;  many believe that the inflated bulge on the outer disc rim that is caused by the collision between the gas stream and the outer disc is responsible \citep[see e.g.][]{wh82}.
 The X-ray modulation  of 4U\thinspace 1916$-$053 shows striking  variability \citep{sm88}. These variations repeat on a $\sim$4 day period, and are caused by the precession of the accretion disc \citep[see e.g.][]{chou01}. When \citet{ret02} made power density spectra (PDS) of the X-ray lightcurve of 4U\thinspace 1916-05, they found peaks corresponding to both the X-ray and optical periods. After removing the dipping intervals, the optical peak was removed; hence the dips were shown to occur on the superhump period. \citet{ret02} concluded that that the observed superhumps arose from the same thickened region of the outer disc that caused the absorption of the X-rays, allowing superhumps to be seen in high inclination LMXBs. 
 
\begin{table}
\centering
\caption{Journal of XMM-Newton observations of the {M31} core. A1--A3 are available in the public archive, while P4--P6 are proprietary observations.  Two observations were made during 2004 July 19; the first observation (a) started at 01:42:12, and the second (b) started at 13:11:22. }\label{journ}
\begin{tabular}{lllllll}
\noalign{\smallskip}
\hline
\noalign{\smallskip}

Observation & Date &  Exp  & Filter\\
\noalign{\smallskip}
\hline
\noalign{\smallskip}
A1 &  2000 Jun 25 &  34 ks& Medium \\
A2 &  2001 Jun 29&  56 ks &Medium \\
A3 &  2002 Jan 6 &  61 ks&  Thin\\
P4 &  2004 Jul 17 &   18 ks & Medium\\
P5 &  2004 Jul 19a & 22 ks & Medium\\
P6 &  2004 Jul 19b & 27 ks & Medium\\
\noalign{\smallskip}
\hline
\noalign{\smallskip}

\end{tabular}
\end{table}

\section{Observations and data analysis}
\label{obs}

In addition to the three XMM-Newton observations analysed by \citet{tru02}, we conducted a programme of four $\sim$20 ks observations over 2004, July 16--19. We present results from our analysis of the archival data, along with three of the four 2004 observations; the other observation suffered from flaring in the particle background over 90\% of the observation  and is not considered further here. A journal of observations is presented in Table~\ref{journ}.

We analysed data from the pn, MOS1 and MOS2 instruments, which share the same 30$\arcsec$$\times$$30\arcsec$ field of view.  We used version 6.0.0 of the $SAS$ software suite\footnote[1]{\texttt http://xmm.vilspa.esa.es} to obtain the data products, as well as the latest calibration data. 
For each observation, we selected  a  circular extraction region with a 40$\arcsec$ radius, centred on Bo 158, and an equivalent source-free region for the background. The background region was on the same chip as the source, and at a similar angular offset from the optical axis.  We extracted lightcurves from the source and background regions in the 0.3--10 keV, 0.3--2.5 keV and 2.5--10 keV energy bands with 2.6 s binning; these were analysed using  $FTOOLS$ version 5.3.1. We also obtained pn spectra of the source and background regions in 4096 channels of 5 eV, and generated response matrix and ancillary response files from the source  spectral files. The spectra were then grouped for a minimum of 50 counts per bin. Spectral analysis was performed using $XSPEC$ 11.3.1.

\section{Results}
\label{res}

\begin{figure}
\resizebox{\hsize}{!}{\includegraphics{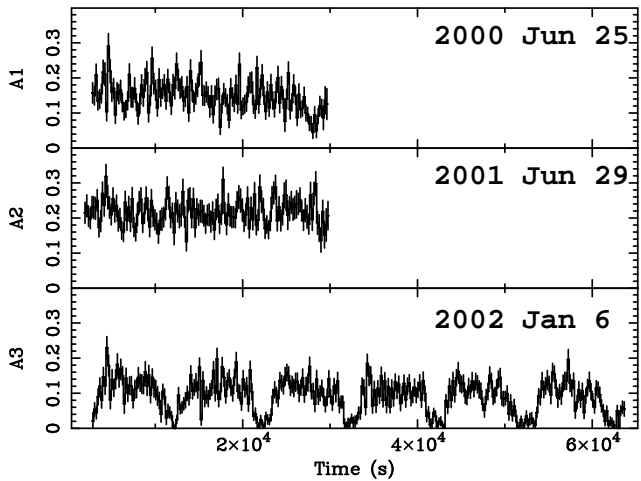}}
\caption{Combined EPIC 0.3--10 keV lightcurves  (in count s$^{-1}$) of Bo 158 from the archival XMM-Newton observations, A1--A3; the dates of each observation are shown. }\label{arclcs}
\end{figure}

The 0.3--10 keV EPIC (MOS1 + MOS2 + pn) lightcurves of Observations A1--A3 are presented in Fig.~\ref{arclcs}, with x- and y- axes set to the same scale, and with 200 s binning. Most striking is Observation A3, with six dipping intervals on a 10017$\pm$50 second period \citep{tru02}; the structure of the dipping is seen to vary substantially even over one observation of 60 ks.
 We emphasise that the 30\% dipping reported  by \citet{tru02} for observation A1 is an overall average; a deep  dip is seen at $\sim$28 ks into the observation, but very little evidence of dipping is observed in the intervals of expected dipping at  $\sim$8 ks and  $\sim$18 ks into the observation. In this regard, Observation A1 resembles the  1985, October lightcurve of 4U\thinspace 1916$-$053, observed by EXOSAT, where deep dips were observed only after the first four orbital cycles \citep{sm88}. Little evidence of variability is seen in the lightcurve of Observation A2. 

\begin{figure}
\resizebox{\hsize}{!}{\includegraphics[angle=0]{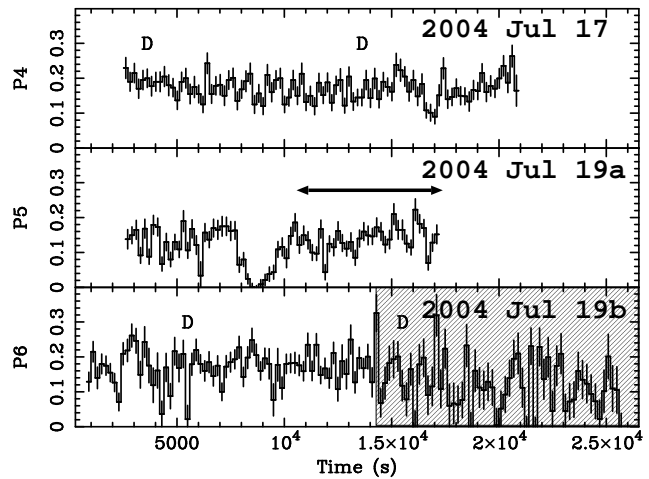}}
\caption{Combined EPIC 0.3--10 keV lightcurves  (in count s$^{-1}$) of Bo 158 from the 2004 XMM-Newton observations, P4--P6; the dates of each observation are shown. The horizontal line in the P5 lightcurve indicates the interval used for spectral fitting. The shaded area of the P6 lightcurve indicates a period of background flaring. Times when dipping is expected  are labelled ``D'', using the deepest part of the dip in P5 as time zero, and the period of \citet{tru02}.}\label{plcs}
\end{figure}

In Fig.~\ref{plcs}, we present the 0.3--10 keV EPIC  lightcurves of Observations P4--P6;  the x-axis is scaled to P6, and the y-axis matches that of Fig.~\ref{arclcs}. The most prominent feature is the dip in P5; it has a depth of $\sim$100\% and a  total duration of $\sim$2500 s. Using the  period  of \citet{tru02}, we identified the expected times of dipping, labelled `D', in  P4 and P6, using the deepest part of the dip in P5 as phase zero. 

In the P4 lightcurve, there is no evidence for dipping during the  first expected dip  interval, but some evidence of dipping $\sim$4000 s after the second interval, 20 cycles away from the dip in P5. Hence, this possible dip   would require a period that is either $\sim$200 s shorter or $\sim$320 s longer than the 10017 s given by \citet{tru02}.  However, there is no other evidence for these other periods in the lightcurves of P4, P5 or P6; hence it is clear that the dipping behaviour of Bo 158 evolves on a time scale of a few days, just like that of 4U\thinspace 1916$-$053. 

Several emission models were fitted to the 0.3--10 keV pn spectrum of P4, each suffering absorption by material in the line of sight. P4 was chosen because it had the longest interval of persistent emission that was not contaminated by background flares;  the resulting source spectrum contained $\sim$2200 counts. We applied the two models that  \citet{tru02} used to model the spectra of A1--A3, namely a power law model and a Comptonisation model ({\sc comptt} in XSPEC). We also applied  a two component model. The emission of many Galactic LMXBs has been succesfully described by a model consisting of a blackbody and a cut-off  power law \citep[e.g.][]{cbc95, mjc98, bcb03}; we approximate this model to a blackbody + power law model, because of the narrow pass-band. Table~\ref{spec} shows the best fits to the spectrum with each model;  uncertainties are quoted at the 90\% confidence level.

\begin{figure}
\resizebox{\hsize}{!}{\includegraphics[angle=270]{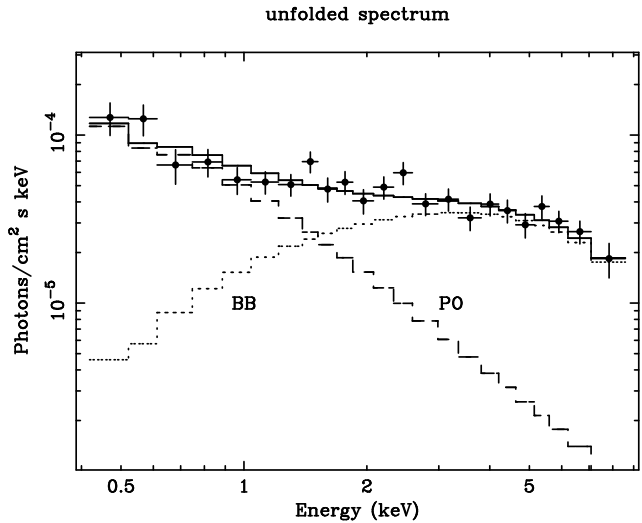}}
  \caption{ Unfolded EPIC-pn SED of Bo 158 from observation P4, described by the best fit two component model. The dotted line represents the blackbody component (BB), while the dashed line shows the power law component (PO). The solid line shows the sum of the two components. The blackbody component dominates the flux above 1.5 keV.}\label{sed}
\end{figure}

We find that the best fit parameters for the power law and {\sc comptt} models agree well with the values presented by \citet{tru02}; however, the two-component model provided the best fit.
 We present the unfolded SED for the two component fit in Fig.~\ref{sed}; the emission from the blackbody (BB) and power law (PO) components are shown separately. We see that the blackbody dominates the emission above 1.5 keV.
We find a 0.3--10 keV flux of $\sim$2$\times 10^{-12}$ erg cm$^{-2}$ s$^{-1}$ for all the  fits to the  P4 data; this gives a 0.3--10 keV luminosity of $\sim$1.4$\times 10^{38}$ erg s$^{-1}$, assuming a distance of 760 kpc \citep{vdb00}.  

\begin{table*}
\centering
\caption{Best fits to the P4 (2004 July 16) pn spectrum of Bo 158 for several emission models; these are power law ({\sc po}), Comptonisation ({\sc comptt}),  and blackbody + power law ({\sc bb+po}) models, suffering line-of-sight absorption by the interstellar medium. $N_{\rm H}$ is the equivalent hydrogen column density of the absorber, $T_0$ is the seed photon temperature ({\sc comptt}), $T$ is the blackbody temperature ({\sc comptt, bb+po}), $\tau$ is the optical depth ({\sc comptt}) and $\Gamma$ is the photon index ({\sc po, bb+po}). The goodness of fit is given by $\chi^2$/dof, and $F_{\rm 0.3-10}$ is the unabsorbed flux in the 0.3--10 keV band. }\label{spec}
\begin{tabular}{llllllllll}
\noalign{\smallskip}
\hline
\noalign{\smallskip}

Model & $N_{\rm H}$ & k$T_{\rm 0}$ & k$T$  & $\tau$ & $\Gamma$ & $\chi^2$/dof & $F_{0.3-10}$ \\
 & atom cm$^{-2}$ & keV & keV & & & & 10$^{-12}$ erg cm$^{2}$ s$^{-1}$\\
\noalign{\smallskip}
\hline
\noalign{\smallskip}
PO & 0.1 & $\dots$ & $\dots$ & $\dots$ & 0.57$\pm$0.09 & 30/24 & 2.3$\pm$0.3\\
COMPTT & 0.1 & 0.08$\pm$0.07 & 2.4$\pm$0.6 & 17$\pm$2 & $\dots$  & 22/19 & 2.2$^{+0.2}_{-2.0}$\\
BB+PO & 0.1 & $\dots$ & 2.0$\pm$0.2 & $\dots$ & 2.0$\pm$0.3 & 19/19 & 2.1$\pm$0.8\\
\noalign{\smallskip}
\hline
\noalign{\smallskip}

\end{tabular}
\end{table*}

We then extracted the SED from the interval of $\sim$persistent emission in P5 indicated in Fig.~3 by a horizontal line. The resulting background-subtracted SED contained 452 counts in the 0.3--10 keV band, which we divided into 10 spectral bins.  The spectral shape of the P5 SED was consistent with that of P4, with a luminosity of 1.2$\pm$0.2$\times$10$^{38}$ erg s$^{-1}$.  Hence the luminosity of persistent emission in P5 is consistent with that of P4 within 3$\sigma$.

 The depth of dipping in A1 varies from $\sim$0 to $\sim$70\%  with no significant change in the mean intensity,  suggesting that the amplitude of dipping is not simply anticorrelated with the source luminosity.  Instead, the variation in dipping behaviour  may be caused by disc precession.  This hypothesis motivated our simulation of the accretion disc in Bo\thinspace 158, using three dimensional smoothed partical hydrodynamics, discussed in Sect.~\ref{sph}.

The lightcurves of XMM-Newton observations of Bo 158 show no eclipses; hence, we know that we are not viewing the system edge on. If the disc were tilted with respect to the binary plane, and precessing, then one might expect to observe dips in some part of the disc precession cycle, but not in others.  Dipping is observed throughout observation A3; this suggests that the dipping phase in the disc precession cycle lasts $\ga$60 ks. The A1 lightcurve covered three intervals of expected dipping, yet only one dip is seen, toward the end of the observation; we suggest that this dip signals the onset of the dipping phase. Contrariwise, P5 and P6  appear to sample the end of the dipping phase, as a dip is observed in P5, yet no dips are seen in P6.

\section{SPH simulation of the disc}
\label{sph}

\subsection{Binary Parameters}
The accretion disc was modelled using a three-dimensional Smoothed Particle Hydrodynamics (SPH) computer code that has been described in detail in \citet{mur96,mur98}, \citet{truss00}, and  \citet*{fhm05}. We assumed the orbital period to be 10017 s, as obtained by \citet{tru02}. 
 Table 3 lists the system parameters adopted. 

 The dipping source Bo 158 is a bright globular cluster X-ray source, with a 2.78 hr binary period. Thirteen Galactic globular clusters contain bright X-ray sources; twelve of these are neutron star LMXBs, while the primary of the other one is unknown \citep[see e.g.][]{int04}. Hence Bo 158 is a likely neutron star LMXB, and we assume the primary mass to be 1.4 M$_{\odot}$.

For the secondary, we considered a main sequence star and a white dwarf, since 4U\thinspace 1916$-$053 has a likely white dwarf secondary \citep[e.g.][]{chou01}. For a Roche lobe-filling main sequence star the approximate relation m $\simeq$ 0.11 $P_{\rm hr}$ holds \citep[e.g.][]{fkr92}, giving a mass of $\sim$0.30 M$_{\odot}$. If instead the star is a white dwarf, using the  mass
radius relation of  \citet{nau72} and Roche geometry gives an implausibly
small mass of 0.005 M$_{\odot}$. 
 Assuming a main sequence secondary, we found the mass ratio to be  0.2, indicating that superhumps and disc precession were likely. 

Finally, the luminosity of the system was taken to be 1.4$\times$10$^{38}$ erg s$^{-1}$, the 0.3--10 keV luminosity of Bo 158 in Observation P4. Such a high luminosity may be expected to cause warping of the accretion disc, even for a previously flat disc \citep[see e.g.][]{pring96}; warping is discussed in Sect. 5.3.

The accretion disc had an open inner boundary condition in the form of a hole of radius $r_{1}=0.025a$,  where $a$  is the binary separation, centred on the position of the primary object. Particles entering the hole were removed from the simulation. Particles that re-entered the secondary Roche lobe were also removed from the simulation as were particles that were ejected from the disc at a distance $>0.9a$ from the centre of mass.

We assumed an isothermal equation of state and that the  dissipated energy was radiated from the point at which it was generated,  as electromagnetic radiation.  The \citet{ss73} viscosity parameters were set to $\alpha_{low}$ = 0.1 and $\alpha_{high}$ = 1.0, and the viscosity state changed smoothly as described in \citet{truss00}. The SPH smoothing length, $h$, was allowed to vary in both space and time and had a maximum value of $0.01a$. 

\subsubsection{The gas stream}
We simulated the mass loss from the secondary by introducing particles at the inner Lagrangian point ($L_1$). The mass transfer rate and the particle transfer rate were provided as input parameters,  and the mass of each particle was derived from these parameters. A particle was inserted with an initial velocity (in the orbital plane) equal to the local sound speed of the donor, $c_{\rm D}$, in a direction prograde of the binary axis. The z velocity of the inserted particle was chosen from a Gaussian distribution, with a zero mean and a variance of  0.1$c_{\rm D}$. However, the inflation of the site of collision between the gas stream and outer disc was not modelled.

\subsubsection{The initial non-warped accretion disc}

The simulation was started with zero mass in the accretion disc and with the central radiation source switched off. A single particle was injected into the simulation every $0.01\Omega_{orb}^{-1}$ at the $L_1$ point as described above until a quasi-steady mass equilibrium was reached within the disc. This was taken to be when the number of particles inserted at the $L_1$ point, the mass transfer rate, was approximately equal to number of particles leaving the simulation at the accretor, the accretion rate. The simulations were continued for another 3 orbital periods to ensure mass equilibrium. The number of particles in the simulated accretion disc was approximately 40,000 giving a good spatial resolution; the average number of `neighbours', i.e. the average number of particles used in the SPH update equations, was 8.2 particles. The simulated disc encountered the Lindblad 3:1 resonance and  became eccentric. The disc precessed in a prograde direction giving rise to superhumps in the simulated dissipation light-curves \citep[c.f.][]{fhm04}. The radiation source was then turned on which gave rise to a very small number of particles being ejected from the accretion disc. 

\begin{table*}
\label{table:DiscParameters}
\begin{center}
\caption{ Binary system parameters for the system modelled. The columns are:  the total system mass,  the system orbital period, the rate of mass loss from the secondary,  the system mass ratio, the physical luminosity of the central radiation source and  the ratio of the physical luminosity to the Eddington limit  for the system. 
  }
\begin{tabular}{ccccccc}
\noalign{\smallskip}
\hline
\noalign{\smallskip}
Parameter & $M_t$       & $P_{orb}$  & $\dot{M}_{sec}$ & q &  Physical luminosity ($L_{*}$) & $L_{*}/L_{edd}$    \\
     & $M_{\odot}$ & $days$     & $M_{\odot}yr^{-1}$ & ($M_{2}/M_{1}$) &  ($erg\ s^{-1}$) \\
\noalign{\smallskip}
\hline
\noalign{\smallskip}
Value     & 1.8       & $0.1159$ & $2.22\times10^{-8}$ & 0.2  & $1.4\times10^{38}$ & 0.8 \\
\noalign{\smallskip}
\hline
\noalign{\smallskip}
\end{tabular}
\end{center}
\end{table*}

\subsection{Surface finding algorithm \& self-shadowing}
 Accretion-powered radiation from the inner regions of the disc and the accreting object itself exert a force on the irradiated disc surface. Following \citet{pring96} the radiation source is modelled as a point source at the centre of mass of the accretor.
To apply this force, particles on the surface of the accretion disc had to be identified. We used a convex hull algorithm to find the surface particles as described in \citet{mur98} and \citet{fhm05}. A ray-tracing algorithm was used to determine regions of self-shadow. For each particle found on the disc surface a light-ray was projected from the particle to the position of the radiation source at the centre of the disc. The particle was deemed to be illuminated by the radiation source if this light-ray did not intersect any disc material between the particle surface position and the radiation source  (i.e. the particle could see the central radiation source). The radiation force was only applied to particles that were considered to form part of the disc surface and were illuminated by the central radiation source.

\subsection{Disc warping and precession measure}

  For an optically thick disc, a warp can develop  as a result of the radiation force \citep[e.g.][]{pring96,od01,fhm05}. This is due to the fact that  any radiation absorbed by a specific region on the disc surface will be later re-radiated from  the same spot, normal to the disc surface. Hence any anisotropy in the disc structure will cause an uneven distribution of back-reaction forces  on the disc surface, further perturbing the disc. A sufficiently high luminosity can induce and sustain a  warp even in an originally flat disc \citep[][ and references therein]{pring96}.

The two measures defined by \citet{lar97} were used to measure the disc warping and the amount of warp precession. They defined an angle $j$ as the angle between the total disc angular momentum vector and the angular momentum vector for a specific disc annulus, i.e.

\begin{equation}\label{eq:22}
\cos\ j = \frac{\mathbf{J}_A \cdot \mathbf{J}_D}{\big | \mathbf{J}_A \big | \big | \mathbf{J}_D \big |}. 
\end{equation} 

\noindent The term $\mathbf{J}_A$ is the total angular momentum within the specific annulus and was calculated by summing the angular momentum for each particle within the annulus. The term $\mathbf{J}_D$ is the total disc angular momentum and was calculated by summing all the angular momenta for all particles within the disc. An angle $\Pi$ was also defined which measures the amount of precession of the disc angular momentum relative to the initial binary orbital angular momentum, $\mathbf{J}_O$

\begin{equation}\label{eq:23}
\cos\ \Pi = \frac{\left(\mathbf{J}_O \times \mathbf{J}_D\right) \cdot \mathbf{u}}
{\big | \mathbf{J}_O \times \mathbf{J}_D \big | \big | \mathbf{u} \big |}
\end{equation} 

\noindent where $\mathbf{u}$ is any arbitrary vector in the binary orbital plane.

\subsection{Numerical  modelling results} 
\label{NUMERICAL RESULTS}

Prior to irradiation, the disc  was asymmetric about the binary axis and precessed in a prograde direction relative to the inertial frame. After switching the radiation source on we ran the model for a further 50 orbital periods,  and found that this illumination introduced a warp in the disc. When the radiation source was removed,  the warp would dissipate and the disc would return to the orbital plane.

\begin{figure}
\resizebox{\hsize}{!}{\includegraphics{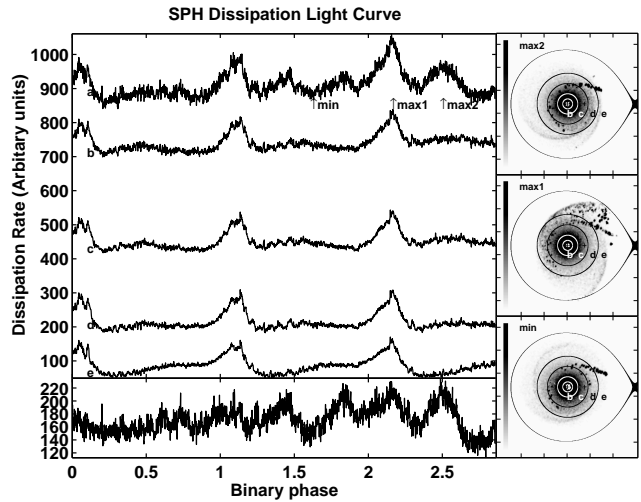}}
  \caption{ The top left plot shows SPH viscous dissipation light curves for different regions of the accretion disc. The top curve, labelled $a$, is for the whole accretion disc, then each curve in descending order corresponds to dissipation  in the disc at radii $> 0.05a (b)$, $> 0.1a (c)$, $> 0.2a (d)$ and $> 0.3a (e)$ respectively. The light curve minimum and two maximum points are indicated. The lower left-hand plot is the signal from the disc inner region and was generated by subtracting light curve $b$ from light curve $a$. The plots on the right-hand side are disc dissipation maps that correspond to the light curve minimum and two maxima. The circles labelled $b$, $c$, $d$ and $e$ show the distances from the primarily that correspond to the light curves $b$, $c$, $d$, $e$ respectively.}
  \label{fig5}
\end{figure}

 As a result of the disc precession, viscous stresses in the disc vary significantly with time. Figure~\ref{fig5} shows how the resultant energy dissipation in different regions of the disc varies with time. The disc luminosity was not modelled in detail. We assume that the luminosity was directly related to the disc regions with significant energy release through viscous dissipation. The viscous dissipation heats the gas in the accretion disc and it is assumed that the heat is radiated away from the point at which it was generated. Superposed on a steady signal there is a repeating series of ``humps." The spacing of the humps corresponds to the superhump period. The humps consist of three separate major components. The maximum occurs at phase $\sim$0.2, followed by  a  secondary maximum at phase $\sim$0.5, and the minimum at $\sim$0.7;  however the relative strengths of the humps vary from cycle to cycle. We stress that the dissipation lightcurves are only a diagnostic for the disc properties, i.e. for identifying the the superhumps, as well as the intervals when the disc structure is most extended or most compressed. They are not representative of  true optical lightcurves.

In order to determine the superhump period, $P_{\rm sh}$, we obtained a power density spectrum from  $\sim$30 superhump cycles of the simulated light curve. We estimated the superhump period to be $(1.035\pm0.005)P_{\rm orb}$.  This implies the precession period of the outer regions, $P_{\rm prec} = (29 \pm 1) P_{\rm orb}$, or 81$\pm$3 hr.

\begin{figure*}
  \resizebox{\hsize}{!}{\includegraphics{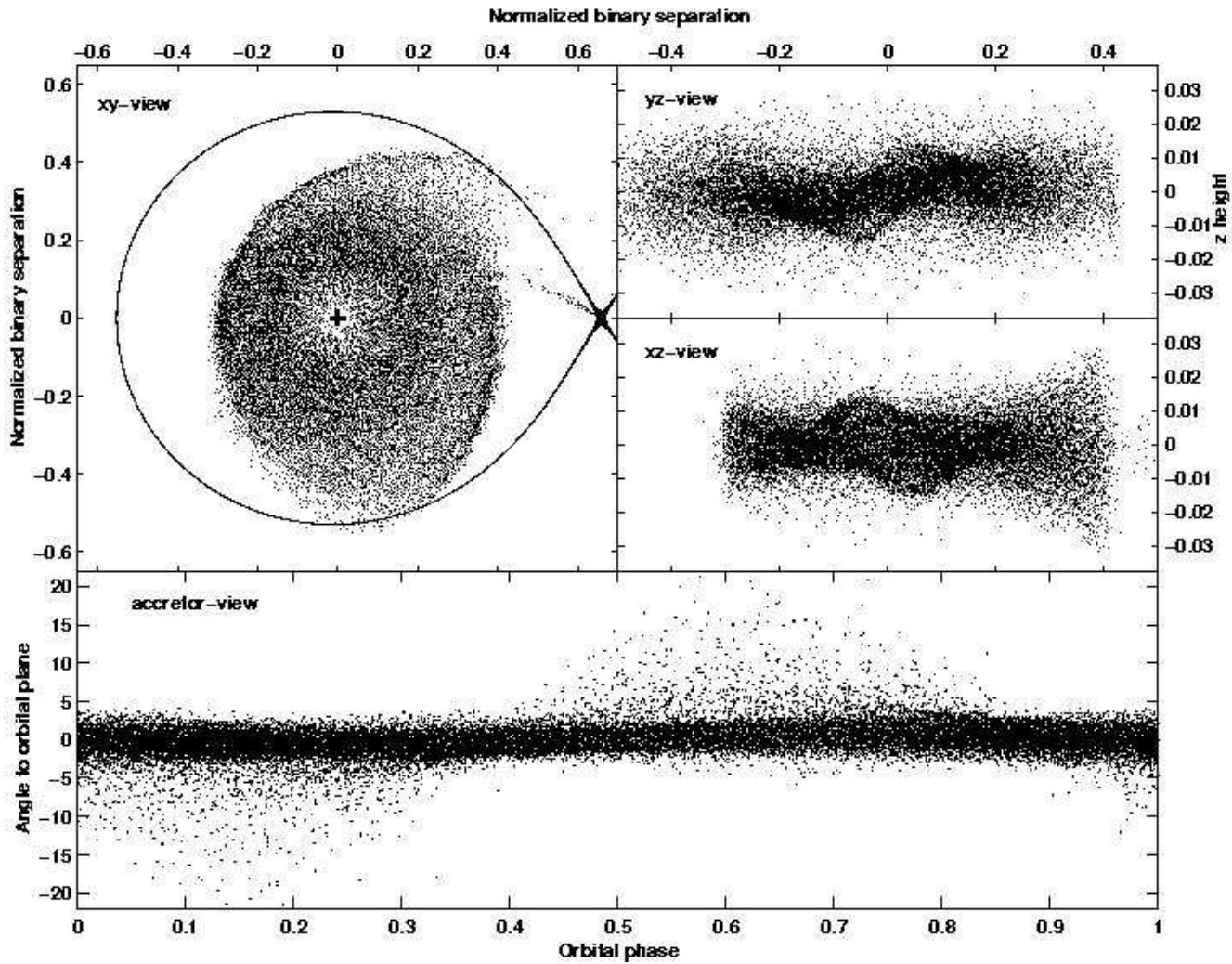}}
  \caption{ Particle projection plots for the SPH model. The position of each particle is indicated by a small black dot. The plot labelled  xy-view is a plan view of the accretion disc as seen from above the disc. The cross at the centre of the plot shows the position of the primary object. The solid dark line is the Roche lobe of the primary and the $L_1$ point is to the right and middle of the plot. The two plots xz-view and yz-view are particle projection plots on a plane perpendicular to the orbital plane and through the system axis. The bottom plot, accretor-view, shows the particle distribution as seen from the compact object. The horizontal axis is the orbital phase, the $L_{1}$ point is at phase 0 and the stream/disc impact region is at approximately phase 0.9. The vertical axis is the angle, in degrees, between a particle and the orbital plane when viewed from the compact object. The disc material flows from right to left.}
  \label{figure:figure_1}
\end{figure*}

Fig. \ref{figure:figure_1} contains projection plots for the time period corresponding to the peak labelled ``max1'' in Fig.~\ref{fig5}, at phase $\sim$2.2.
 A very strong spiral density compression wave can be seen at the upper edge of the disc. This wave is so intense that it is removing material from the accretion disc and returning it back to the Roche lobe of the secondary, see \citet{fhm04} for a full detailed description of a similar system with a mass ratio of $0.1$.

The two upper right-hand plots of Fig. \ref{figure:figure_1}, labelled yz-view and xz-view, are side views of the disc in the y-z and x-z directions respectively. The yz-view plot is a projection view of the disc as seen from the secondary, similarly the xz-view is a projection plot with the secondary located to the right of the plot.  The disc warp is clearly apparent in these two plots. The warp is odd symmetrical about the centre of the disc. The maximum value of the warp is located at a distance approximately $0.1a$ either side of the primary position, see yz-view of Fig. \ref{figure:figure_1}.   

The lower plot of Fig. \ref{figure:figure_1}, labelled accretor-view, shows the distribution of the particles as seen from the compact object. The horizontal axis is the binary orbital phase, $L_1$ is located at phase $0$ and disc material flows from right to left with the stream-disc impact region located at approximately phase $0.9$. The vertical axis is the elevation angle of the particle as seen from the primary position. From this plot it can be seen that the radiation force has pushed disc material out of the orbital plane. The warp reaches a maximum height  above the orbital plane at phases $\sim$0.1 and $\sim$0.6,  and has minima at phases $\sim$0.45 and $\sim$0.95. We also see that the disc remains mainly in the orbital plane, although it can be seen that there is a small S-wave in the structure of the disc.

\begin{figure}
    \resizebox{\hsize}{!}{\includegraphics{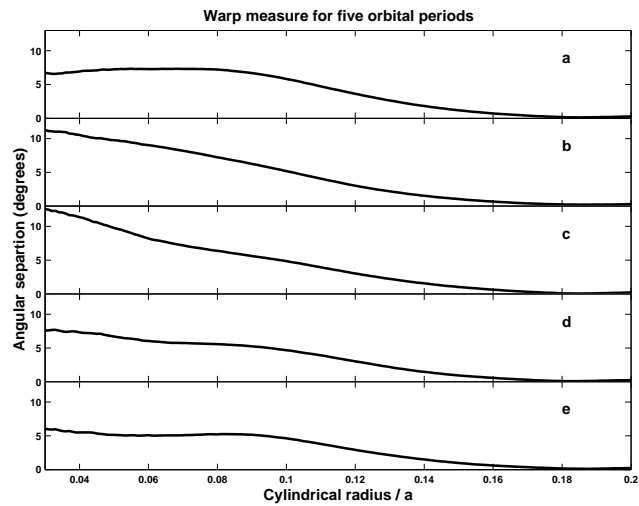}}
  \caption{ Radial warp profiles evaluated using equation (\ref{eq:22}). The vertical axis is the warp amplitude. The horizontal axis is distance from the primary object normalized such that the binary separation is $1$. Plots (a), (b), (c), (d) and (e) are for five consecutive orbital cycles.}
  \label{figure:profiles}
\end{figure}

The warp amplitude and size precessed as a solid body in a retrograde direction relative to the inertial frame.
The warp precession period, $P_{\rm warp}$, was determined by applying equation (\ref{eq:23}) to each time step of the simulation. This gave a precession angle relative to some arbitrary start angle for each simulation time step. The warp precession rate was then determined by fitting a straight line to  these data using a Numerical Recipes least squares method, \citep{press86}. The precession rate was found using the gradient of the line extracted from the least squares fit. We found that  $P_{\rm warp}$ $\sim$11 P$_{\rm orb}$.
  Figure~\ref{figure:profiles} shows the radial profile of the warp for five consecutive orbital cycles; the maximum extent of the warp is 6--11$^{\circ}$ above the plane of the disc.

\section{Discussion}
\label{discuss}

 We report  a single $\sim$100\% dip in the  0.3--10 keV EPIC lightcurve of Bo\thinspace 158 from Observation P5, but find little evidence of dipping in Observations P4 and P6.
\citet{tru02} propose two models to explain the energy independence of dipping  that they inferred from Bo\thinspace 158: (1) absorption by a highly ionised region and (2) partial covering of an extended source by an opaque absorber that occults varying fractions of the source. However, in their figure showing both non-dip and dip SEDs, the two SEDs converge at high energies, showing that the dipping is indeed energy dependent.

 The dipping behaviour of  many Galactic LMXBs has been well described by a single model: absorption of a point-like blackbody plus progressive covering of an extended emission region by an extended absorber \citep{mjc97}. This extended emission is caused by unsaturated inverse-Comptonisation of cool photons on hot electrons in an accretion disc corona that has a radius of $\sim$10,000--500,000 km \citep{cbc04}. The saturated dipping exhibited by 4U\thinspace 1624$-$490 during a 1985 EXOSAT observation  is particularly strong evidence for an extended absorber \citep{cbc95}. The 0.1--200 keV observations of 4U\thinspace 1916$-$053 with Beppo-SAX have shown that the absorber is so dense that $\sim$100\% photo-electric absorption occurs up to 10 keV; in fact, the dipping is seen up to 40 keV \citep{mjc98}. However, the high luminosity of Bo 158, together with the large blackbody contribution, mean that the dipping is unlikely to be energy-independent.

Disc precession is inferred from the 0.3--10 keV lightcurves of Bo 158, as is expected given its extreme mass ratio (short orbital period). As such, it resembles the Galactic superhumping LMXB 4U\thinspace 1916$-$053. Since the LMXB Bo 158 is in a globular cluster near the centre of M31, it is unlikely that the optical period will ever be known.  However, our Fourier analysis of the simulated dissipation lightcurves indicates a superhump period that is 3.5$\pm$0.5\% longer than the orbital period.  Given the association between the dips and superhump period reported by \citet{ret02}, the 10017 s period may be the superhump period, in which case, the orbital period would be $\sim$4\% shorter. Such shortening of the period would not dramatically affect the outcome of our SPH modelling.

 Our simulations of the disc show two distinct types of variability in the disc structure. First is the elongation and prograde precession of the disc due to tidal interactions with the secondary at the 3:1 resonance; the disc precesses on  period of 81$\pm$3 hr. We also see warping of the accretion disc, driven by irradiation of the disc surface by the central X-ray source; the warp is stable exhibits retrograde precession on a  $\sim$31-hr period.  

It is therefore important to establish which region is responsible for the observed variation in dip morphology. The lightcurves of observations A1--P6 show no eclipses. From Kepler's law and the ratio of the secondary ratio to the binary separation \citep{egg83}, the secondary has an angular radius of $\sim$15$^{\circ}$; hence the inclination $\la$75$^{\circ}$. We see from Fig.~6 that the disc warp does not deviate from the plane of the disc by more than 11$^{\circ}$ in our simulations, suggesting that the observed dips are likely to evolve on the disc precession period.

\section{Conclusions}

We have analysed three new XMM-Newton observations of the M31 dipping LMXB Bo 158, in addition to re-analysing the three observations discussed in \citet{tru02}. The newer observations spanned $\sim$3 days in 2004, July. We find that that the relationship between source intensity and depth of dipping is not simple, as described by \citet{tru02}. Instead, we believe that the observed variation in dipping behaviour is caused by precession in the accretion disc; dipping would be confined to a limited phase range in the disc precession cycle.

We modelled the accretion disc with 3D SPH, and found prograde disc precession on a 81$\pm$3 hr period, as well as radiatively driven disc warp that precessed on a ~31 hr period in a retrograde fashion. We find that the disc precession is most likely to affect the observed dipping behaviour.
Hence, we predict that the dipping behaviour of Bo\thinspace 158 experiences a 81$\pm$3 hour cycle; this period is consistent with the observed variation of the dips.

\section*{Acknowledgments}

 We would like to thank the anonymous referee for their constructive comments. R.B. gratefully acknowledges support from PPARC.

\bibliographystyle{mn2e}
\bibliography{mnrasm31}

\end{document}